\documentclass{ws-ijmpd}
\usepackage[super,compress]{cite}
\begin{document}

\markboth{M.M.Lapola, Seetesh Prande}
{Wormholes in energy-momentum-conserved  $f(R,T)$ Gravity}

\catchline{}{}{}{}{}
%
%%%%%%%%%%%%%%%%%%%%%%%%%%%%%%%%%%%%%%%%%%%%%%%%%%%%%%%%%%%%%%%%%%%%

\title{Traversable wormholes in conservative $f(R,T)$  gravity: an effective energy-momentum tensor approach}

\author{{M. M. Lapola$^{*\ddagger}$}, {Seetesh Prande$^{\ddagger}$}}

\address{{$^*$} Núcleo Comum de Engenharia, Centro Universitário Fundação Hermínio Ometto, Av. Dr. Maximiliano Baruto, 500 - Jardim Universitario, Araras - SP, 13607-339, Brazil}

\address{$^\ddagger$ Turing Enterprises, Inc., 548 Market Street, PMB 18282, San Francisco, CA 94104}

\maketitle

%\begin{history}
%\received{Day Month Year}
%\revised{Day Month Year}
%\end{history}

\begin{abstract}
{We present a wormhole construction in the framework of the $f(R,T)$ extended gravity theory, with $R$ being the Ricci curvature scalar and $T$ the trace of the energy-momentum tensor, in the case in which the energy-momentum tensor $T$ is conserved. From the conservation equation we find that the $f(R,T)$ function is linear in $T$. By developing the field equations, we find a traversable wormhole that for a range of the model parameter values, has no need to be filled by exotic matter.} 
\end{abstract}


\ccode{PACS numbers:}

%\tableofcontents

\section{Introduction}	

\label{intro}

Wormholes are shortcuts in space-time that connect two distant regions in the universe. The wormholes theory was proposed for the first time by Einstein and Rosen in 1935 \cite{key-1}, when they established a solution based on the Schwarzschild metric, that describes bridges between two different areas of space-time modeled with the Einstein's General Relativity vacuum field equations. After a period of almost three decades of numbness in the area, in 1962, Wheeler and Fuller published a paper showing that the Einstein-Rosen bridges are unstable and would collapse as soon as they are formed, preventing even light from being able to cross them \cite{key-2}.

The revival of the wormhole theory occurred in 1988 with Morris and Thorne, when they found a metric for a ``traversable'' wormhole, i.e. a wormhole whose material content keeps its throat open \cite{key-3}. Solutions to Einstein's field equations for humanly traversable wormholes, as shown by Morris and Thorne, violate the null energy condition. Matter violating the null energy condition is named {\it exotic matter}.

The fluid that permeates the inner of a wormhole must be anisotropic, i.e., with an energy-momentum tensor that, in addition to the energy density in its temporal component, has radial and tangential pressures, the latter in its angular components.

Some alternative theories of gravity have been capable of describing non-exotic matter in the wormhole interior, such as the Gauss-Bonnet theory \cite{key-5}.

In the realm of extended gravity, an extension of the well-known $f(R)$ theories of gravity \cite{key-4}, for which $R$ is the Ricci scalar, was proposed by Harko and collaborators by further inserting a dependence on $T$, the trace of the energy-momentum tensor, denoting the $f(R,T)$ gravity \cite{Harko/2011}. The theory has been successfully tested in many areas \cite{key-15,key-16,key-17,key-18,key-19,key-20,Debarata/2019}.

In the present article, starting from the metric of a static wormhole, we will calculate the field equations using the $f(R,T)$ gravity formalism, but imposing the conservation of the energy-momentum tensor. It is well-known that the $f(R,T)$ formalism predicts the non-conservation of the energy-momentum tensor \cite{Harko/2011}. Harko himself gave a thermodynamic interpretation, explicitly stating that it describes an irreversible matter creation processes \cite{harko/2014}. The particle creation would correspond to an irreversible energy flow from the gravitational field to the created matter constituents. Prior to this article, some mechanism to evade the non-conservation of the energy-momentum tensor were presented in the cosmological scenario \cite{alvarenga/2013,chakraborty/2013}. Later, this mechanism was taken to the compact objects hydrostatic equilibrium application \cite{dos_santos_jr/2019,carvalho/2020}.

It is interesting to quote that although so far wormholes have not been detected, attempts to do so have been constantly proposed \cite{shaikh/2018,tsukamoto/2012,kuhfittig/2014,nandi/2017}.

This work is organized as follows: in Section 2, we present the $f(R,T)$ gravity theory, including the energy-momentum tensor conservation equation. In Section 3 we show the wormhole metric and energy-momentum tensor. In Section 4 we detail the energy conditions that must be obeyed for the wormhole to be traversable in our model. The results to combine the conservative $f(R,T)$ gravity with a linear equations of state (EoS) for the radial and tangential pressures are presented in Section 5. In Section 6 we show the non-exotic conditions for the wormhole to be traversable. Section 7 is dedicated to show that our results are remarkably in line with the results of a previous work \cite{moraes2017modeling}, in the framework of a non-conservative $f(R, T)$ gravity. The discussions of our results are in Section 8. Finally, the deduction of the covariant derivative of the energy-momentum tensor is shown in details in Appendix A.

\section{The $f(R,T)$ gravity theory}
\label{sec:frt}

In the $f(R,T)$ gravitational theory, the action reads \cite{Harko/2011}

\begin{equation}\label{frt1}
S=\int\left[\frac{f(R,T)}{16\pi}+\mathcal{L}_{m}\right]\sqrt{-g}d^{4}x,
\end{equation}
with $g$ being the determinant of the metric, $f(R,T)$ a general function of the argument and natural units are assumed. In our case, we will take the matter Lagrangian density $\mathcal{L}_{m}$ to be the matter-energy density $\rho$.

Varying this action with respect to the metric components $g_{\mu\nu}$, we obtain the general form of the field equations as

\begin{equation}\label{frt2}
   f_{R}(R,T)R_{\mu\nu}-\frac{1}{2}f(R,T)g_{\mu\nu}+\left(g_{\mu\nu}\Box-\nabla_{\mu}\nabla_{\nu}\right)f_{R}(R,T)=8\pi T_{\mu\nu}+f_{T}(R,T)\left(T_{\mu\nu}-\rho g_{\mu\nu}\right),
\end{equation}
in which $f_{R}(R,T)=\partial f/\partial R$, $R_{\mu\nu}$ is the Ricci tensor, $T_{\mu\nu}$ is the energy-momentum tensor and $f_{T}(R,T)=\partial f/\partial T$. It is clear the dependence of the theory on the choice for $\mathcal{L}_{m}$. An alternative to evade that question was proposed \cite{moraes/2019}.

The general expression for the covariant derivative of the energy-momentum tensor is given by\cite{Barrientos/2014}

\begin{equation}
\nabla^{\mu}T_{\mu\nu}=\frac{f_{T}(R,T)}{8\pi-f_{T}(R,T)}\left[\left(T_{\mu\nu}+\rho g_{\mu\nu}\right)\nabla^{\mu}\ln f_{T}(R,T)+\nabla^{\mu}\rho g_{\mu\nu}-\frac{1}{2}g_{\mu\nu}\nabla^{\mu}T\right].
\end{equation}

\section{Wormhole metric and energy-momentum tensor}

The static traversable wormholes, as shown by Morris and Thorne 
\cite{key-3}, in Einstein's General Relativity, violate
the weak, strong and dominant energy conditions, i.e., somewhere
near the throat of the wormhole, someone must be able to encounter
some negative energy density. Assuming that this traversable wormholes
are time independent, non-rotating and spherically symmetric, in Schwarzschild coordinates, without loss
of generality, the spacetime metric of the wormhole is of the form 

\begin{equation}
ds^{2}=e^{2\varphi(r)}dt^{2}-\left[\frac{1}{1-\frac{b(r)}{r}}\right]dr^{2}-r^{2}\left[d\theta^{2}+sin^{2}\theta d\phi^{2}\right],
\end{equation}
where $\varphi(r)$ is the redshift function, $b(r)$ is the shape function and $r=r_{0}$ is the throat of the wormhole, the minimum value that $r$ can assume. In order for the spacetime geometry to tend to the appropriate flat limit, $\underset{r\rightarrow\infty}{lim}\varphi(r)=\varphi_0$, with constant $\varphi_0$. 

The energy-momentum tensor is given by

\begin{equation}
T_{\nu}^{\mu}=diag\left(\rho,-p_{r},-p_{t},-p_{t}\right),
\end{equation}
with $\rho$ being the density and $p_{r}$ and $p_{t}$ being the radial and tangential pressures, respectively. 

\section{Energy Conditions}

As mentioned in the Introduction, wormholes in General Relativity lead to a violation of causality with the mathematical prediction of the occurrence of exotic matter inside them, i.e., matter that violates the energy conditions \cite{key-3}. They are described as the following: 

$\bullet$ The strong energy condition (SEC) says that gravity should be always attractive. In terms of the energy-momentum tensor components, it reads $\rho+ \Sigma_j p_j\geq0$, $\forall j$; 

$\bullet$ The dominant energy condition (DEC) is an indication that the velocity of energy transfer cannot be higher than the speed of light, which leads to $\rho\geq|p_j|$, $\forall j$;

$\bullet$ The weak energy condition (WEC) shows that the energy density measured by any observer should be always non-negative, i.e., $\rho\geq0$ and $\rho + p_j\geq0$, $\forall j$. 

$\bullet$ The null energy condition (NEC) is a minimum requirement from SEC and WEC, i.e. $\rho+p_j\geq0$, $\forall j$. The violation of NEC implies that the above energy conditions are not validated. 

\section{Wormhole field equations for the conservative $f(R,T)$ gravity}

Let us assume the general form $f(R,T) = R+h(T)$, with $h(T)$ being a function depending only on the trace of the energy-momentum tensor. In this way, if $h(T) = 0$ we recover the Einstein-Hilbert action for general relativity in action (1). In this case, from the formalism presented in Section 2, $f_{R}(R,T)=1$, and $f_T(R,T)=h_T$. 

By doing the left-side of Eq.(3) equal to zero, it reads

%\begin{equation}
%\left[\left(T_{\mu\nu}+\rho g_{\mu\nu}\right)\nabla^{\mu}ln(h_{T})+\nabla^{\mu}\rho g_{\mu\nu}-\frac{1}{2}g_{\mu\nu}\nabla^{\mu}T\right]=0, 
%\end{equation}

%which in its mixed form results

\begin{equation}
\left[\left(T_{\nu}^{\mu}+\rho\delta_{\nu}^{\mu}\right)\nabla^{\mu}\ln(h_{T})+\nabla^{\mu}\left(\rho-\frac{1}{2}T\right)\delta_{\nu}^{\mu}\right]=0.
\end{equation}
This is the mechanism used in the literature when trying to obtain conservative versions of the $f(R,T)$ gravity \cite{alvarenga/2013,chakraborty/2013,dos_santos_jr/2019,carvalho/2020}, i.e., to force $\nabla^\mu T_{\mu\nu}=0$ and to find a solution to $h_T$, and consequently, $h(T)$.

Writing, from (5), the trace of the energy-momentum tensor as $T=\rho-p_{r}-2p_{t}$, Eq.(6) reads, for $\mu=\nu=r$, 

\begin{equation}
\left[\left(p_{r}+\rho\right)\frac{\partial}{\partial r}ln\left(h_{T}\right)+\frac{1}{2}\frac{\partial}{\partial r}\left(\rho+p_{r}+2p_{t}\right)\right]=0, 
\end{equation}

We need to solve this differential equation for $h_T$. A viable way is to invoke an EoS. Using the relations below between pressures and density,

\begin{equation}
p_{r}=\beta\rho,
\end{equation}

\begin{equation}
p_{t}=\gamma\rho,
\end{equation}
with $\beta$ and $\gamma$ being constants, then, the conservation equation (7) becomes

\begin{equation}
\left(1+\beta\right)\rho\frac{\partial\ln(h_{T})}{\partial{r}}+\frac{1}{2}\left(1+\beta+2\gamma\right)\frac{\partial \rho}{\partial{r}}=0.
\end{equation}

Since $T$ is linear in $\rho$, we can rewrite (10) as

\begin{equation}
\left[\frac{\partial \ln(h_{T})}{\partial T}+\frac{1}{T}\frac{\left(1+\beta+2\gamma\right)}{2(1+\beta)}\right]\frac{\partial T}{\partial r}=0,
\end{equation}

which leads to

\begin{equation}
\frac{T}{h_{T}}\frac{\partial h_{T}}{\partial T}=-\frac{\left(1+\beta+2\gamma\right)}{2(1+\beta)}.
\end{equation}

Then

\begin{equation}
T\frac{\partial h_{T}}{\partial T}=\alpha h_{T},
\end{equation}

where

\begin{equation}
-\frac{\left(1+\beta+2\gamma\right)}{2(1+\beta)}\equiv\alpha.
\end{equation}

If $h_{T}=\lambda T^{\alpha}$, i.e, a power law expression, then

\begin{equation}
\frac{dh_{T}}{dT}=\alpha T^{\alpha-1}=\alpha\frac{h_{T}}{T}.
\end{equation}

Hence, the $f(R,T)$ function that satisfies the conservation law of $T^{\mu\nu}$, for a linear EoS, is given by

\begin{equation}
h(T)=\frac{\lambda T^{-\frac{1+\beta-2\gamma}{2(1+\beta)}}}{\frac{1+\beta-2\gamma}{2(1+\beta)}},
\end{equation}
from which we can define

\begin{equation}
\tilde{\lambda}\equiv\frac{1+\beta-2\gamma}{2(1+\beta)},
\end{equation}
and write

\begin{equation}
h(T)=\frac{\lambda}{\tilde{\lambda}}T^{-\tilde{\lambda}},
\end{equation}
so the derivative with respect to $T$ reads

\begin{equation}
\frac{dh(T)}{dT}=-\lambda T^{-1-\tilde{\lambda}}.
\end{equation}

Now, the non null components of the field equation (2) are 

\begin{equation}
-\left(\frac{r_0}{r^2}\right)^2=8\pi\rho+\frac{1}{2}\frac{\lambda}{\tilde{\lambda}}T^{-\tilde{\lambda}},
\end{equation}

\begin{equation}
\left(\frac{r_0}{r^2}\right)^2=-8\pi p_r+\lambda T^{-\tilde{\lambda}}\left[\frac{1}{2\tilde{\lambda}}+T^{-1}(\rho+p_r)\right],
\end{equation}

\begin{equation}
-\left(\frac{r_0}{r^2}\right)^2=-8\pi p_t+\lambda T^{-\tilde{\lambda}}\left[\frac{1}{2\tilde{\lambda}}+T^{-1}(\rho+p_t)\right],
\end{equation}
where the shape function is defined  as in {[}30{]} by
\begin{equation}
b(r)=r_{o}\left(\frac{r_{o}}{r}\right).
\end{equation} % continuar daqui
This shape function obeys the following conditions
{[}12-14{]}

\begin{equation}
b(r)<r,
\end{equation}

\begin{equation}
1-\frac{b(r)}{r}\geq0,
\end{equation}

\begin{equation}
b^{'}(r)<\frac{b(r)}{r},
\end{equation}

\begin{equation}
\underset{r\rightarrow\infty}{lim}\frac{b(r)}{r}=0
\end{equation}

Since, by the equation of motion $p_t=-\rho$ with this above ansatz for $b(r)$, so

\begin{equation}
\gamma=-1
\end{equation}

In terms of the asymmetry we have 

\begin{equation}
\Delta=p_t-p_r=(-1-\beta)\rho
\end{equation}

Using too the field equations (23), (24) and (25):

\begin{equation}
\Delta(r)=-\left(1+\beta\right)\rho=\frac{\frac{2b}{r^{3}}}{8\pi+f_{T}(R,T)},
\end{equation}

so, fro the asymmetry be proportional to the energy density $\rho$

\begin{equation}
\frac{1+\beta}{2}=\frac{\left(1+\frac{f(R,T)}{8\pi\rho2}\right)}{\left(1+\frac{f_{T}(R,T)}{8\pi}\right)}
\end{equation}

then

\begin{equation}
f_{T}=\frac{f(R,T)}{2\rho}.
\end{equation}

From Eq. (35)

\begin{equation}
\frac{1+\beta}{2}=1,
\end{equation}

then
\begin{equation}
\beta=1
\end{equation}

Since the trace is $t=2\rho$, and $f(R,T)=R+h(T)$, the Eq. (37) implies that

\begin{equation}
h(T)=\frac{h(T)}{T}
\end{equation}

Using Eq. (22) with $\beta=1$ and $\gamma=-1$ we obtain exactly the same solution and, as a consequence $h_{T}(T)=\lambda$ is a constant.
Thus, we have proved that the only expression for $f(R,T)$ that obeys the conservation of energy-momentum tensor, with linear dependence in $\rho$ for the radial and tangential pressures, is a function that is linear in the trace $T$ of the energy-momentum.

\section{Non-exotic conditions for the wormhole to be traversable }

Summarizing the results, for the solution $\beta=1$, and $\gamma=-1,$
and then f(R,T) linear with T, we have

\begin{equation}
p_{r}=\rho,
\end{equation}

\begin{equation}
p_{t}=-\rho,
\end{equation}

\begin{equation}
T=2\rho=-\Delta(r).
\end{equation}

Then, substituting these results in the field equation (24) leads to an expression for the energy density given by

\begin{equation}
\rho(r)=\frac{-r_{o}}{\left(8\pi+\lambda\right)r^{4}}.
\end{equation}

Then, for $\rho(r)\geq0\Rightarrow\lambda\ < -8\pi,$i.e, the Dominant Energy Condition (DEC) can be respected at the wormhole's throat ($r=r_0$)

And

\begin{equation}
(\rho+p_{r})=2\rho=\frac{-2r_{0}}{\left(8\pi+\lambda\right)r^{4}},
\end{equation}

\begin{equation}
(\rho+p_{t})=0.
\end{equation}

\begin{figure}[h]
\centering
\includegraphics[scale=0.9]{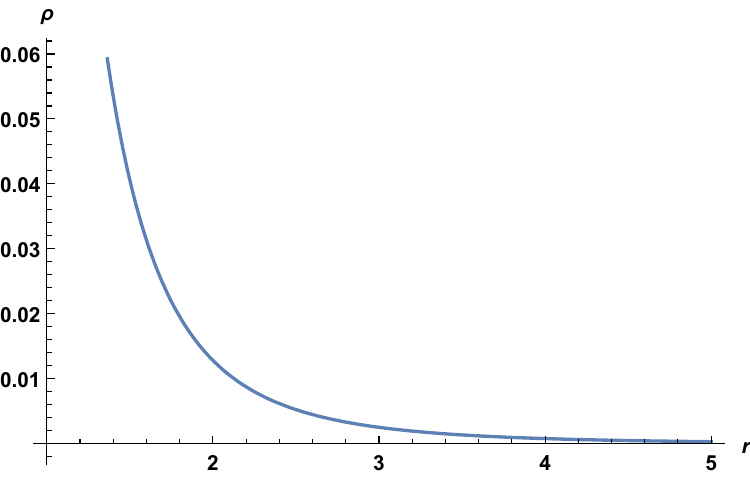}
\caption{The energy density versus $r$ for $r_0=1$ and $\lambda=-30$.}
\label{fig:density-profile}
\end{figure}

\begin{figure}[h]
\centering
\includegraphics[scale=1]{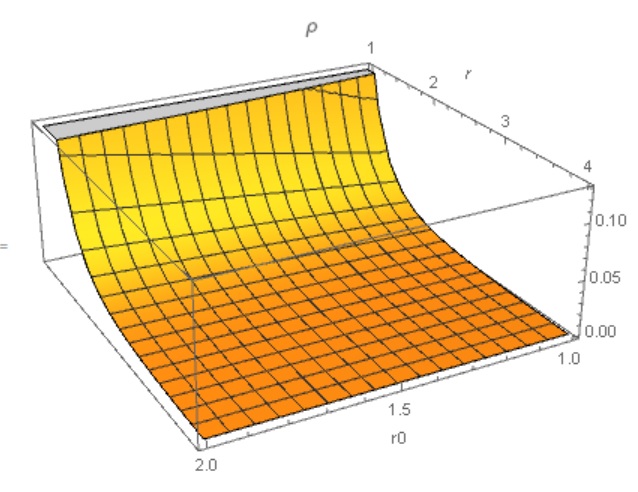}
\caption{The energy density $\rho$ profile with $r_0=1$ to $2$, $r=1$ to $4$ and $\lambda=-30$.}
\label{fig:density-3d}
\end{figure}

\begin{figure}[h]
\centering
\includegraphics[scale=0.9]{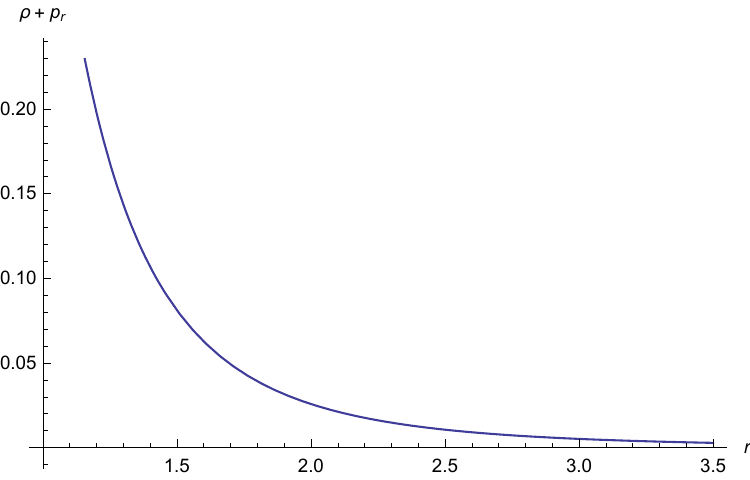}
\caption {The null energy condition (NEC) is obeyed for  $r_0=1$ and $\lambda=-30$.}
\label{fig:rho-pr}
\end{figure}

So, for these solutions, as seen in Figures (1), (2) in the limit $\lambda < -8\pi$, the Null Energy Condition (NEC), are also respected, no need to fill the wormhole with exotic matter, i.e, $p_r=\rho$, as we found above in the Eq. (40). Using only General Relativity framework, as in the Morris-Thorne work \cite{key-3}, they found that $-p_r>\rho$, leading to the exotic matter solution as the condition to maintain the throat of the wormhole opened.
Finally, the Strong Energy Condition (SEC) are still obeyed for linear conservative $f(R,T)$ model:

Strong energy condition (SEC), is also obeyed for $\lambda < -8\pi$, with the solutions given by Eqs. (40) and (41)

\begin{equation}
\rho+\sum_{j}p_{j}=\rho+p_{r}+p_{t}=\rho>0,
\end{equation}

\section{More general form for shape function as solution of the field equations}

With the linear function $f(R,T)=R+\lambda T$, the field equations (23), (24), and (25) admits general solutions to $\rho$, $p_r$ and $p_t$, given by

\begin{equation}
\rho=\frac{b'}{\left(8\pi+\lambda\right)r^{2}},
\end{equation}

\begin{equation}
p_{r}=-\frac{b}{\left(8\pi+\lambda\right)r^{3}},
\end{equation}

\begin{equation}
p_{t}=\frac{b-b'r}{\left(8\pi+\lambda\right)2r^{3}},
\end{equation}

as was also found in \cite{moraes2017modeling} . In this article, the authors present the modeling of wormholes in the $f(R,T)$ theory in its more general form, without considering the conservation of the energy-momentum tensor.
Considering the relation between pressures as $p_t=np_r$, with $n$ being an arbitrary constant, substituting it on Eqs (48) and (49), leads to the more general shape function

\begin{equation}
b(r)=Ar^{1+2n},
\end{equation}

with $A$ being an integration constant.

Note that for $n=-1$, and $A=r_0$ we recover the more particular case for shape function assumed as an ansatz, given by Eq. (26), which was used in this present work to deduce the energy density and wormhole pressures in the conserved $f(R, T)$ extended gravity model.
Considering Eq.(49) the factor $n$ must be negative to satisfy the preliminary conditions to $b(r)$ shown in Eqs. (27-30).

For any value of $n$ deriving Eq. (49) with respect to $r$, substituting and combining in the Eqs. (46), (47), and (48), the radial and tangential pressure $p_{r}=\beta\rho$, and $p_{t}=\gamma\rho$, respectively, can be written with $\beta$ and $\gamma$ as a function of $n$, as

\begin{equation}
\beta=-\frac{1}{1+2n},
\end{equation}

\begin{equation}
\gamma=-\frac{n}{1+2n},
\end{equation}

where $p_{t}=np_{r},$with $n=\gamma/\beta$. Then, the trace of energy-momentum
tensor is

\begin{equation}
T=\rho-p_{r}-2p_{t}=\left(1+\frac{1}{1+2n}+\frac{2n}{1+2n}\right)\rho.
\end{equation}

Thus,

\begin{equation}
T=2\rho,
\end{equation}

does not depend on $n$ and is exactly the expression we obtained
before, for $n=-1$.

Taking into account the Eq. (8) for the conservation of the energy-momentum tensor, repeated below

\begin{equation}
\left[\left(T_{\nu}^{\mu}+\rho\delta_{\nu}^{\mu}\right)\nabla^{\mu}ln(h_{T})+\nabla^{\mu}\left(\rho-\frac{1}{2}T\right)\delta_{\nu}^{\mu}\right]=0,
\end{equation}

and for $T=2\rho$ (Eq. (54)), it is clear that $h(T)=\lambda T$, and $h_{T}=\lambda$, is the solution of the above equation. Thus, the linear function in $T$, $f(R, T)=R+ \lambda T$ is the unique form for wormholes in the energy-momentum conservative $f(R,T)$ theory, for radial and tangential pressures depending linearly in the energy-density $\rho$ with the general solution to $b(r)$ for traversable wormholes expressed by Eq. (49), with $n<0$.

\section{Discussions}
\label{sec:6}

We have constructed, in the present work, a static traversable wormhole within the $f(R, T)$ theory in its conservative form, i.e., assuming the conservation of the energy-momentum tensor.
For the conservation equation to be satisfied, using linear EoS for the radial and tangential pressures ($p_r$ and $p_t$) inside the wormhole, we find a $f(R, T)$ linear in the trace of the energy-momentum tensor, as one sees in Eq. (38).
In this way, $p_r$ has the highest pressure value that can be (stiffest fluid there is) and $p_t=-1$, which suggests that can be a dark matter halo surrounding the wormhole, to be investigated in future work.
Then, one obtained for $p_r$ the casual limit, i.e, the sound speed in the wormhole interior is given by

\begin{equation}
\frac{dp_r}{d\rho}=1=c,
\end{equation}

where $c$ is the speed of light.
We found a fluid that is not exotic in the same way as obtained in similar wormhole work in linear f(R, T) gravity. \cite {rosa2022non}. 
In this cited work the analyzed traversable wormhole solutions in the linear form of $ f\left (R,
T\right)= R+\lambda T $ gravity satisfying the energy conditions, but without taking into account the energy-momentum tensor conservation. 
As we know in its original model, the theory of gravitation $f(R,T)$ does not conserve the energy-momentum tensor, which is very far from Einstein's general relativity, in which the conservation of the energy-momentum tensor is a condition. fundamental that leads to the field equations.
The novelty of our work is to assume conservation of the energy-momentum tensor, thus being in greater agreement with reality.

Unlike a Morris-Thorne wormhole, in which the exoticity of matter is shown with $p_r<-\rho$, using General Relativity, we find $p_r=\rho$, denoting an ultra-relativistic (non-exotic) matter and whose radial pressure is positive,

As we have mentioned in the previous sections all the conditions for the shape function $b(r)$, established for the present traversable wormholes, are respected by the ansatz that we adopted (see Eqs. (27 to 31)).
Furthermore, we have assumed the redshift function is a constant ($\varphi=const.$), implying in null tidal gravitational forces through the wormhole, i.e. a static wormhole.

It is important to remark that to obtain wormhole solutions satisfying the WEC is not a trivial task. Here, the extra degrees of freedom provided by $f(R,T)$ gravity formalism allowed the material solutions to obey WEC so that the referred wormhole can be said to be filled by non-exotic matter, recalling that Morris and Thorne defined ``exotic matter'' as matter violating WEC.
In our case, we find a wormhole that obbey all the energy conditions in the limit that $\lambda < -8\pi$.

\appendix
\section{Covariant Derivative expression for $T_{\mu\nu}$}

In this paper we use the covariant derivative equation of energy-momentum tensor on $f(R,T)$ gravity as one see in \cite {Barrientos/2014} 

To deduce the covariant derivative expression for energy-momentum tensor in Eq. (3), we use the following mathematical identity 

\begin{equation}
\left(\Box\nabla_{\nu}-\nabla_{\nu}\Box\right)\phi\equiv R_{\mu\nu}\nabla^{\mu}\phi,
\end{equation}

which is valid for any field $\phi$. In our case, the filed can be
substituted by $f(R,T)$. Aplying the covariant derivative in the
left side of field equation (Eq. 5), and use this identity above.

\begin{equation}
\nabla^{\mu}\left[f_{R}(R,T)R_{\mu\nu}-\frac{1}{2}f(R,T)g_{\mu\nu}+\left(g_{\mu\nu}\Box-\nabla_{\mu}\nabla_{\nu}\right)f_{R}(R,T)\right]
\end{equation}

\begin{equation}
=\left(\nabla^{\mu}f_{R}(R,T)\right)R_{\mu\nu}+f_{R}(R,T)\nabla^{\mu}R_{\mu\nu}-\frac{1}{2}\nabla^{\mu}f(R,T)g_{\mu\nu}+\nabla^{\mu}\left(g_{\mu\nu}\Box-\nabla_{\mu}\nabla_{\nu}\right)f_{R}(R,T)
\end{equation}

\begin{equation}
=\nabla^{\mu}R_{\mu\nu}-\frac{1}{2}\nabla^{\mu}f(R,T)g_{\mu\nu}+{\left(R_{\mu\nu}\nabla^{\mu}f_{R}(R,T)\right).}
\end{equation}

Assuming since the begining that $f_{R}(R,T)=1$, then $\nabla^{\mu}(f_{R}(R,T))=0$,
and $\nabla^{\mu}g_{\mu\nu}=0$, as in general relativity framework.

Finally, using the covariant derivative of $f(R,T)$ in general form,
given by

\begin{equation}
\nabla^{\mu}f(R,T)=f_{R}(R,T)\nabla^{\mu}R+f_{T}(R,T)\nabla^{\mu}T,
\end{equation}

where $f_{R}(R,T)$=1.

The we have the final expression for the covariant derivative of Eq.
5 left side

\[
\nabla^{\mu}\left[f_{R}(R,T)R_{\mu\nu}-\frac{1}{2}f(R,T)g_{\mu\nu}+\left(g_{\mu\nu}\Box -\nabla_{\mu}\nabla_{\nu}\right)f_{R}(R,T)\right]
\]

\[
=\nabla^{\mu}R_{\mu\nu}-\frac{1}{2}\nabla^{\mu}Rg_{\mu\nu}-\frac{1}{2}g_{\mu\nu}f_{T}(R,T)\nabla^{\mu}T
\]

\begin{equation}
=-\frac{1}{2}g_{\mu\nu}f_{T}(R,T)\nabla^{\mu}T.
\end{equation}

In the last step above we apply the Bianchi identity for the Einstein
tensor, $\nabla^{\mu}\left(R_{\mu\nu}-\frac{1}{2}Rg_{\mu\nu}\right)=\nabla^{\mu}G_{\mu\nu}=0$.

This non-vanishing term on the lhs covariant derivative of field equations
plays an essential role in $f(R,T)$ gravity.

Then we can write the covariant derivative of field equation (5) as

\begin{equation}
\nabla^{\mu}\left[8\pi T_{\mu\nu}-f_{T}(R,T)T_{\mu\nu}-f_{T}(R,T)\rho g_{\mu\nu}\right]+\frac{1}{2}g_{\mu\nu}f_{T}(R,T)\nabla^{\mu}T=0.
\end{equation}

Isolating $\nabla^{\mu}T_{\mu\nu}$, and apllying the $\nabla^{\mu}$
operator in the other terms results

\begin{equation}
\nabla^{\mu}T_{\mu\nu}=\frac{f_{T}(R,T)}{8\pi-f_{T}(R,T)}\left[\left(T_{\mu\nu}+\rho g_{\mu\nu}\right)\nabla^{\mu}lnf_{T}(R,T)+\nabla^{\mu}\rho g_{\mu\nu}-\frac{1}{2}g_{\mu\nu}\nabla^{\mu}T\right].
\end{equation}

\section*{Acknowledgements}

I would like to express my deep gratitude to Dr. Manuel Malheiro for his invaluable guidance and contributions to this work. This paper is dedicated to his memory, whose passion for science and education inspired all who had the privilege of knowing him.
Thanks for professor Pedro Moraes for the idea to write this article and the review for all calculations doing here.


\begin{thebibliography}{100}

\bibitem{key-1} A. Einstein and N. Rosen, Phys. Rev. 48, 73 (1935)

\bibitem{key-2} R. W. Fuller and J. A. Wheeler, Phys. Rev 128,
919 (1962)

\bibitem{key-3} M.S. Morris and K.S. Thorne, Wormholes in spacetime
and their use for interstellar travel: a tool for teaching general
relativity, Am. J. Phys. 56 (1988) 395

\bibitem{key-5} M.R. Mehdizadeh, M.K. Zangeneh and F.S.N. Lobo, Einstein-Gauss-Bonnet traversable wormholes satisfying the weak energy
condition, Phys. Rev. D 91 (2015) 084004

\bibitem{key-4} T.P. Sotiriou, V. Faraoni, f(R) theories of gravity, Rev. Mod. Phys.82, 451 (2010)

\bibitem{Harko/2011} T. Harko, F.S.N. Lobo, S. Nojiri and S.D. Odintsov, f(R, T) gravity, Phys. Rev. D 84 (2011) 024020

\bibitem{key-15} M.Sharif, I. Nawazish
Annals of Physics 400,37-63 (2019)
	
\bibitem{key-16} E. Elizalde and M. Khurshudyan. Phys. Rev. D 98, 123525 (2018)

\bibitem{key-17} P.V Tretyakov, Cosmology in modified f( R, T)-gravity.
The European Physical Journal C,78, Issue 11, 896, 7 (2018)

\bibitem{key-18} M. Sharif and M. Zubair, Thermodynamics in f(R,T) theory of gravity JCAP, 2012 (2012)

\bibitem{key-19} P. H. R. S.Moraes, J. D. V. Arba{\~{n}}il, G. A. Carvalho, R. V. Lobato, E. Otoniel, R. M. Marinho Jr., M. Malheiro.
Compact Astrophysical Objects in $f(R,T)$ gravity. Invited talk presented at the XIV International Workshop on Hadron Physics, Florian\'opolis, Brazil, March 2018
arXiv:1806.04123v4

\bibitem{key-20} G. A. Carvalho, R. V. Lobato, P. H. R. S. Moraes,  J. D. V. Arba{\~{n}}il, E. Otooniel, R. M. Marinho Jr., M. Malheiro.
Stellar equilibrium configurations of white dwarfs in the f(R, T) gravity. The European Physical Journal C, 77, Issue 12, 871-8 (2017)

\bibitem{Debarata/2019} Debabrata Deb, Sergei V Ketov, S K Maurya, Maxim Khlopov, P H R S Moraes, Saibal Ray, Exploring physical features of anisotropic strange stars beyond standard maximum mass limit in f(R,T) gravity, Monthly Notices of the Royal Astronomical Society, Volume 485, Issue 4, June 2019, Pages 5652 - 5665.

\bibitem{harko/2014} T. Harko, Phys. Rev. D {\bf 90}, 044067 (2014).

\bibitem{alvarenga/2013} F.G. Alvarenga et al., Phys. Rev. D {\bf 87}, 103526 (2013).

\bibitem{chakraborty/2013} S. Chakraborty, Gen. Rel. Grav. {\bf 45}, 2039 (2013).

\bibitem{dos_santos_jr/2019} S.I. dos Santos Jr. et al., Eur. Phys. J. Plus {\bf 134}, 398 (2019).

\bibitem{carvalho/2020} G.A. Carvalho et al., Int. J. Mod. Phys. D {\bf 29}, 2050075 (2020).

%\bibitem{carvalho/2020} Carvalho, G. A., dos Santos Jr, S. I., Moraes, P. H. R. S., Malheiro, M. (2020), International Journal of Modern Physics D, 29(10), 2050075.

%\bibitem{key-8} C. Bambi, A. Cardenas-Avendano, G. J. Olmo and D. Rubiera-Garcia, Phys. Rev. 128, 919 (1962)

%\bibitem{key-9} Lobo, F. S. N. and M. A. Oliveira, Phys. Rev. D 80 (2009) 104012

%\bibitem{key-10}  T. Harko, F.S.N. Lobo, M.K. Mak and S.V. Sushkov,Modified-gravity wormholes without exotic matter, Phys. Rev. D 87
%(2013) 067504

%\bibitem{key-11} P.H.R.S. Moraes, and P. K. Sahoo, Nonexotic matter
%wormholes in a trace of the energy-momentum tensor squared gravity,
%Phys.Rev. D 97, 024007 (2018) 

%\bibitem{key-12} F.S.N. Lobo and M. A. Oliveira, Phys.Rev. D 80,
%104012 (2009)

%\bibitem{key-13} P. Pavlovic and M. Sossich, Eur. Phys. J. C 75,
%117 (2015)

%\bibitem{key-14} M. Visser, Lorentzian Wormholes: From Einstein
%to Hawking (AIP Press, New York, 1995)

\bibitem{shaikh/2018} R. Shaik, Phys. Rev. D {\bf 98} 024044 (2018).

\bibitem{tsukamoto/2012} N. Tsukamoto et al., Phys. Rev. D {\bf 86} 104062 (2012).

\bibitem{kuhfittig/2014} P.K.F. Kuhfittig, Eur. Phys. J. C {\bf 74} 2818 (2014).

\bibitem{nandi/2017} K.K. Nandi et al., Phys. Rev. D {\bf 95} 104011 (2017).

\bibitem{moraes/2019} P.H.R.S. Moraes, Eur. Phys. J. C {\bf 79}, 674 (2019).

\bibitem{Barrientos/2014} José Barrientos O. and Guillermo F. Rubilar Phys. Rev. D 90, 028501 (2014)

\bibitem{Arbanil/2017} Jose D.V. Arbanil and M. Malheiro JCAP11 (2016) 012

\bibitem{Alvarenga/2013} F.G. Alvarenga et al., Phys. Rev. D 87 (2013) 103526.

\bibitem{Elizalde/2013} E. Elizalde, M. Khurshudyan, International Journal of Modern Physics D 28 (15), 1950172

\bibitem{Chakraborty} S. Chakraborty, Gen. Rel. Grav. 45 (2013) 2039.

\bibitem {moraes2017modeling} Moraes, P. H. R. S., and P. K. Sahoo. "Modeling wormholes in f (R, T) gravity." Physical Review D 96.4 (2017): 044038.

\bibitem {rosa2022non} Rosa, João Luís, and Paul Martin Kull. "Non-exotic traversable wormhole solutions in linear $ f\left (R, T\right) $ gravity." arXiv preprint arXiv:2209.12701 (2022).


\end{thebibliography}
 \end{document}